\documentclass[useAMS,usenatbib,usegraphicx]{mn2e}

\title[Circumbinary Planets and Eclipse Timing Variations]
{Prospects of the Detection of Circumbinary Planets With Kepler and CoRoT Using
the Variations of Eclipse Timing}
\author[Schwarz, Haghighipour, Eggl, Pilat-Lohinger \& Funk]
{R. Schwarz $^{1}$\thanks{E-mail:schwarz@astro.univie.ac.at}, 
N.  Haghighipour$^{2}$,  S. Eggl$^{1}$,  E. Pilat-Lohinger$^{1}$ and B. Funk$^{3}$
\\
$^{1}$Institute for Astronomy, University of Vienna, A-1180 Vienna, 
T\"urkenschanzstrasse 17, Austria\\ 
$^{2}$Institute for Astronomy and NASA Astrobiology Institute, 
University of Hawaii, 2680 Woodlawn Dr., Honolulu, HI, 96822 USA\\
$^{3}$Department of Astronomy, E\"otv\"os University, H-1117 Budapest, 
P\'azm\'any P\'eter setany 1/A, Hungary\\
}

\begin{document}

\date{}

\pagerange{\pageref{firstpage}--\pageref{lastpage}} \pubyear{2002}
\maketitle
\label{firstpage}

\begin{abstract}

In close eclipsing binaries, measurements of the variations in binary's eclipse
timing may be used to infer information about the existence of circumbinary
objects. To determine the possibility of the detection of such variations
with CoRoT and {\it Kepler} space telescopes, we have carried out an extensive
study of the dynamics of a binary star system with a circumbinary planet, and calculated 
its eclipse timing variations (ETV) for different values of the mass-ratio and orbital
elements of the binary and the perturbing body. Here, we present the results of
our study and
assess the detectability of the planet by comparing the resulting values of ETVs
with the temporal sensitivity of CoRoT and {\it Kepler}. Results point to
extended regions in the parameter-space where the 
perturbation of a planet may become large enough to create
measurable variations in the eclipse timing of the secondary star.
Many of these variations point to potentially detectable ETVs and the
possible existence of Jovian-type planets.

\end{abstract}


\section{Introduction}

Approximately 70 percent of the main and pre-main sequence stars
are members of binary or multiple star systems. Observational
evidence indicates that many of these systems contain potentially planet-forming
circumstellar or circumbinary disks implying that planet formation
may be a common phenomenon in and around binary stars
\citep{Mathieu94,Akeson98,Rodriguez98,White99,Silbert00,Mathieu00,Trilling07}.

Many efforts have been made to detect such {\it binary-planetary} systems.
In the past two decade, even though single stars were routinely prioritized 
in search for extrasolar planets, many of these efforts were successful and 
resulted in the detection of approximately 40 planet-hosting binary star systems.  
The discovery of these systems have led to the speculation that many more planets may exist 
in and around binaries and prompted astronomers to explore the possibility
of the detection of these planets with different detection methods. 
We refer the reader to {\it Planets in Binary Star Systems} \citep{Hagh10}
for an up to date and comprehensive review of the current state of observational and
theoretical research in this area. 

\begin{figure}
\centerline{
\includegraphics[width=7cm,angle=0]{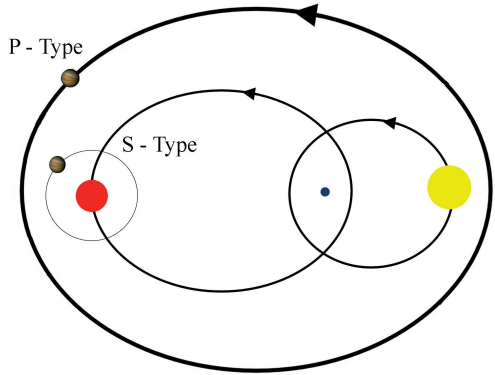}}
\caption{Schematic illustration of S-type and P-type orbits in a binary star 
system. The yellow and red circles represent the stars of the binary revolving
around their common center of mass (the blue circle).}
\label{fig1}
\end{figure}

In general, one can distinguish three types of planetary orbits in 
a binary star system (Fig. \ref{fig1}):

\begin{enumerate}
\item {\bf S-Type,} where the planet orbits one of the two stars,
\item {\bf P-Type,} where the planet orbits the entire binary,
\item {\bf T-Type,} where the planet orbits close to one of the
two equilibrium points $L_4$ and $L_5$ (Trojan planets).
\end{enumerate}

Currently, the most known planets in binary systems are in S-type orbits 
[for more details see e.g., \cite{loh02}, \cite{loh03}, and \cite{haghighi}]. 
The stellar separations of many of these binaries are larger than 100 AU implying
that the perturbation of their farther companions on the formation and dynamical
evolution of planets around their planet-hosting stars may be negligible. However,
in the past few years, radial velocity and Astrometry surveys have been able to identify
five binary star systems with separations of approximately 20 AU where 
one of the stars is host to a Jovian-type planet. These 
systems, namely, GL 86 \citep{Queloz00,Els01}, $\gamma$ Cephei \citep{Hatzes03}, 
HD 41004 \citep{Zucker04,Raghavan06}, HD 196885 \citep{Correia08}, and
HD 176051 \citep{Muterspaugh10} 
present unique cases for the study of planet formation and dynamics in binaries 
as the perturbation of their secondary stars will have significant effects on 
the dynamics of their circumprimary disks and their capability in forming
planets. 

The success of radial velocity and astrometry in detecting planets
in S-type orbits raises the question that whether other techniques can also
detect planets in and around binary stars. A recent success 
has come in the form of the detection of the first P-type planets using the 
ETV method. Modeling the 
variations in the eclipse timing of the binary NN Ser, \citet{beuer} have
shown that this system is host to two planets with minimum masses of 2.2 and 
6.9 Jupiter-masses in a 2:1 mean-motion resonance in circumbinary orbits. 
    
Our work has been motivated by this discovery.
NN Ser (ab) is an eclipsing short period binary which shows long-term ETVs. 
The detection of P-type planets around this system suggests that close, eclipsing 
binaries may be promising candidates in the search for new exoplanets. 
As  many of these binaries lie in the discovery space of CoRoT \citep{goldi} and 
{\it Kepler} \citep{Coughlin10} space telescopes, we focus our study on 
exploring the prospects of the detection of such objects.

The idea of photometric detection of extrasolar planets around eclipsing binaries
was first presented by \cite{schneiderc}. This idea that was later developed by many
authors \citep{schneiderd,doyle,doyle04,deeg08,Muterspaugh10}, is based on the fact that
a circumbinary planet can perturb the orbit of the two stars and create variations in
their eclipse timing. The measurements of these variations, when compared with 
theoretical models, can reveal information about the mass and orbital elements of
the perturbing planet. 

Recently \citet{syb} studied the potential of the ETV method
in detecting giant planets in circular, circumbinary orbits. These authors have shown
that ground-based photometry may have advantage over CoRoT and {\it Kepler} space telescopes 
in detecting circumbinary planets. They suggest that the selective nature of the target 
stars in the fields of view of CoRoT and {\it Kepler} causes the ETV signals
of eclipsing binaries in these regions to be smaller than the detection sensitivities
of these telescopes. 
These authors also suggest that if target binary stars include presumably less stable
contact binaries, the CoRoT's and {\it Kepler}'s capabilities of detecting of 
circumbinary planets increase by four times. 

In this paper, we extend the study by \citet{syb} to binaries with 
planets in eccentric and resonant orbits. It is expected
that similar to the transit timing variation method
where TTV signals are strongly enhanced when the transiting and perturbing planets are
in resonance \citep{agol05,agol07,Steffen05}, a resonant perturbing circumbinary planet 
will also produce high ETV signals.
The goal of our study is to identify regions of the parameter-space for which 
this signal is within the temporal sensitivity of CoRoT and {\it Kepler} space telescopes.

The outline of our paper is as follows. We present our models in section 2 and
discuss their stability in section 3. Section 4 has to do with the calculations
of ETVs and the prospects of the detection of P-type planets. Section 5 concludes 
this study by summarizing the results and discussing their implications.

\section{Model}

We consider a close, eclipsing binary with 
a giant planet in a P-type orbit. To ensure that the effect of the 
variations of the mass-ratio of the binary is included in our study, 
we consider the following three models 
where $m_1$ and $m_2$ are the masses of the primary and secondary stars, 
respectively,
\begin{itemize}
\item {\bf model 1: } $m_1=m_2=0.3\;M_{\rm sun}$,
\item {\bf model 2: } $m_1=1,m_2=0.5\;M_{\rm sun}$,
\item {\bf model 3: } $m_1=m_2=1\;M_{\rm sun}$.
\end{itemize} 
\noindent
 As most of the stars in the solar neighborhood are of spectral type M,
this choice of models ensures that low-mass binary stars are also included in our simulations.

Eclipsing binaries are morphologically classified as
\begin{enumerate}
\item {\bf detached systems,} if neither component fills its Roch lobe (separated stars),
\item {\bf semi-detached systems,} if only one component fills its Roche lobe, and 
\item {\bf over contact systems,} if both components exceed their Roche lobes.
\end{enumerate} 
\noindent
The Roche lobe marks the volume limit at which the star may begin to 
lose substantial amount of matter to its companion. In this study we  
consider a {\bf detached binary} with an initial separation of 0.05 AU. 
For the binary models 1 and 3, this separation corresponds to a period of approximately 
5 and 3 days, respectively. As shown by \cite{goldi}, most candidate eclipsing 
binaries in the discovery space of CoRoT have periods 
between 1 and 10 days. 

When dealing with close binaries such as the models considered here,
the intrinsic eccentricity of the binary ($e_{\rm bin}$) can be neglected.
This is due to the high probability of circularization that is caused by interstellar
tidal forces. In this study, we consider the initial orbital eccentricity of the 
binary to be zero or have very low values. The latter is due to the fact that
the timescale of circularization is dependent upon specific stellar parameters
\citep{zahn}.

\begin{figure*}
\hbox{
\includegraphics[width=4.1cm,angle=270]{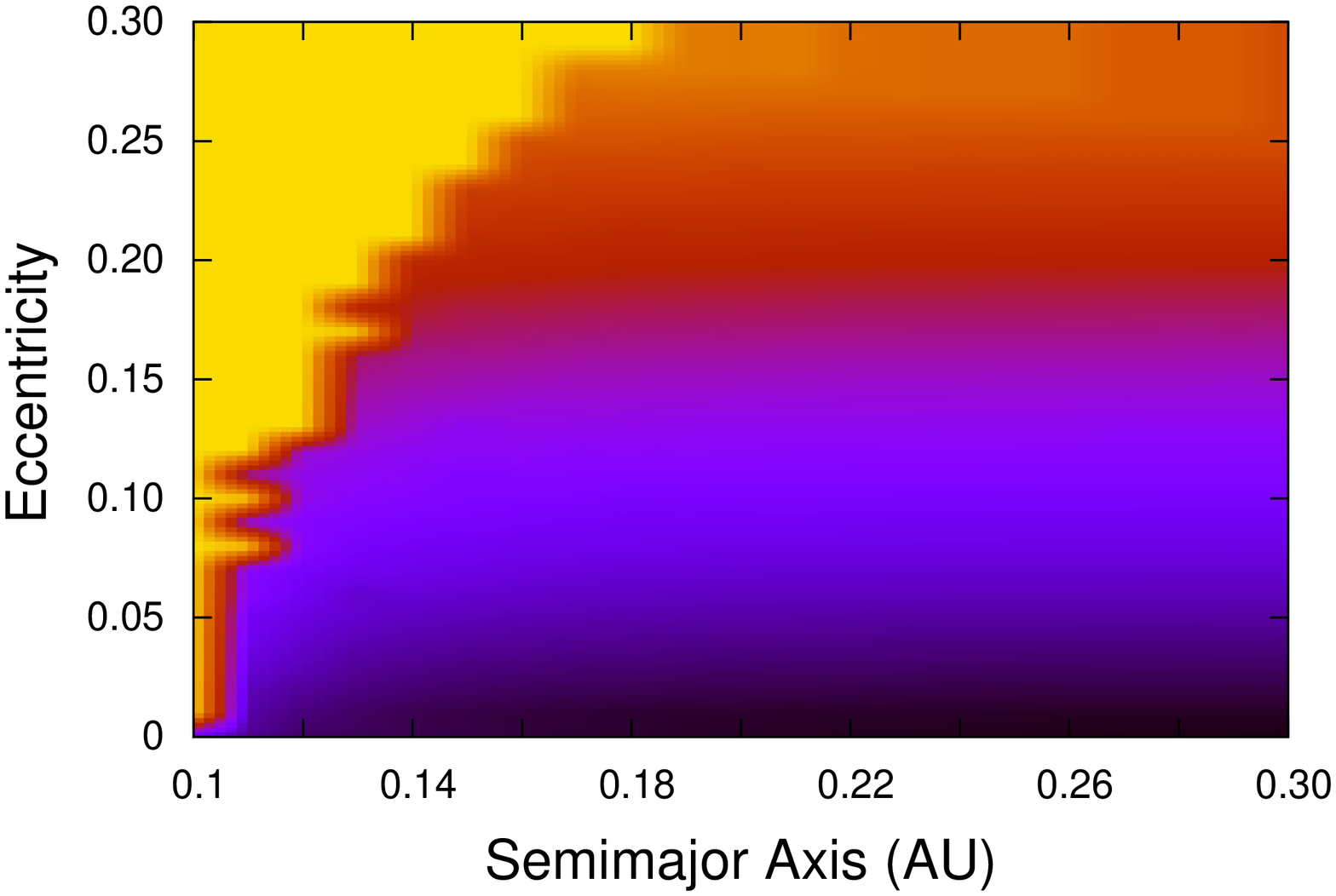}
\includegraphics[width=4.1cm,angle=270]{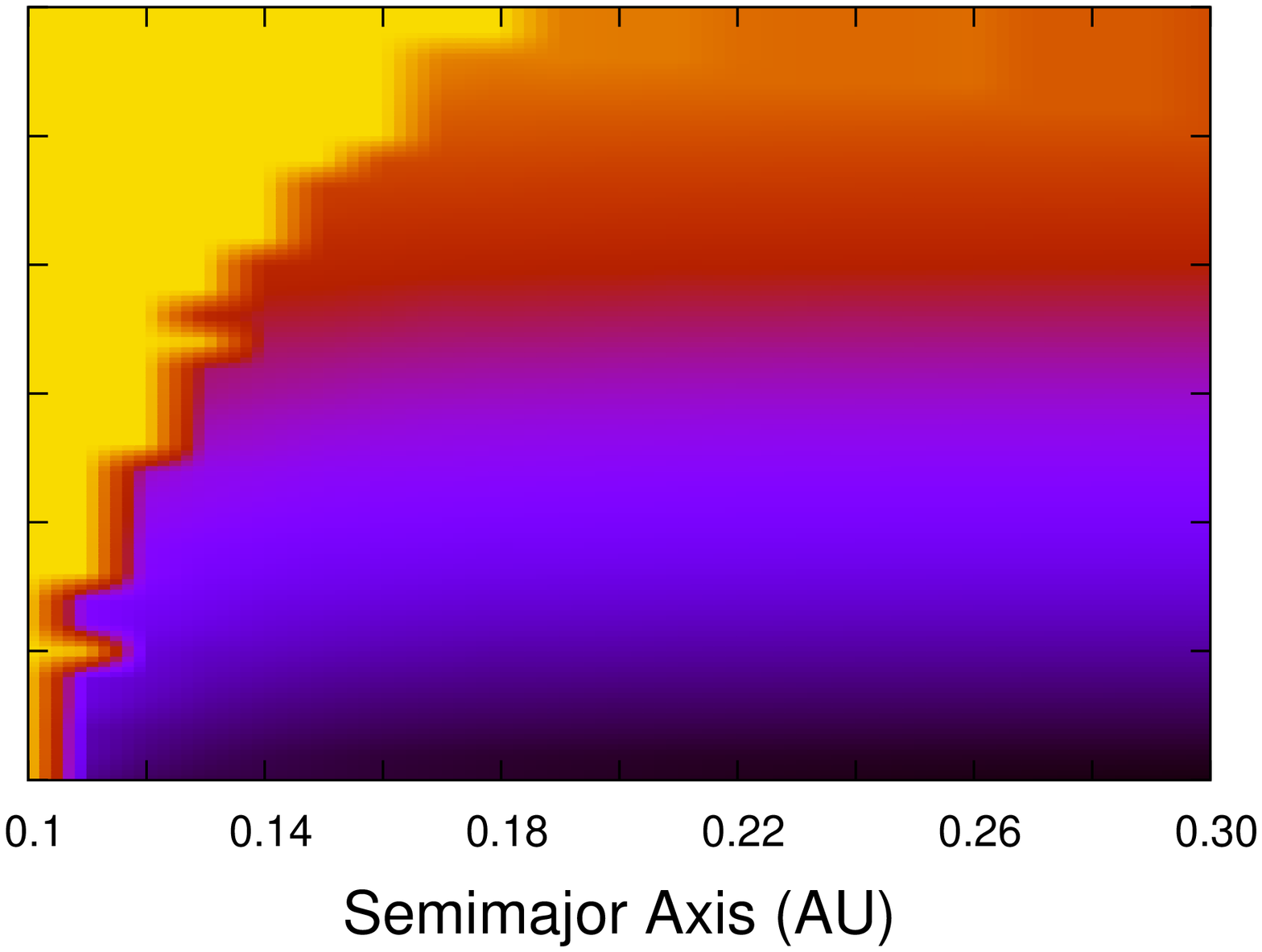}
\includegraphics[width=4.1cm,angle=270]{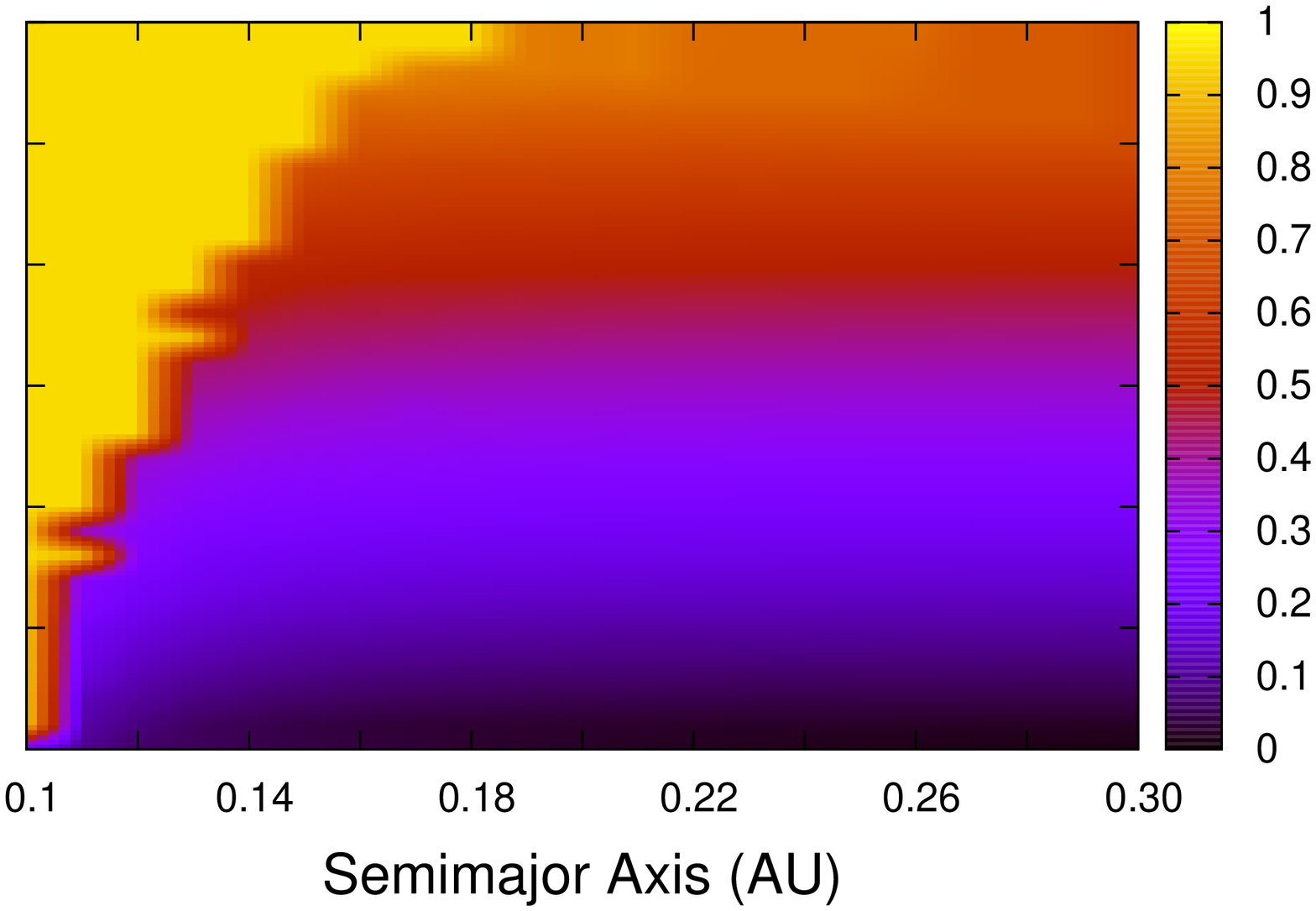}
}
\hbox{
\includegraphics[width=4.1cm,angle=270]{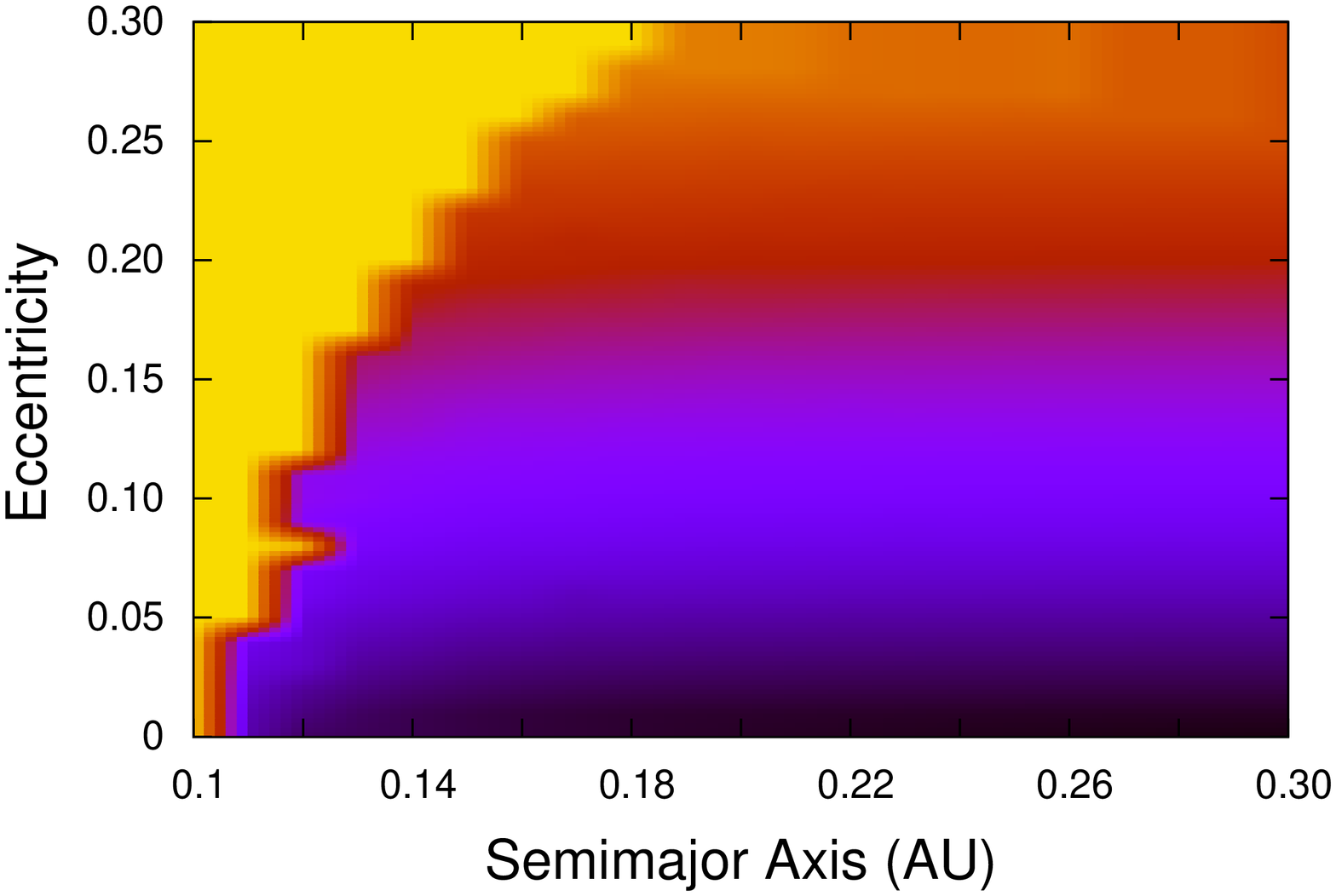}
\includegraphics[width=4.1cm,angle=270]{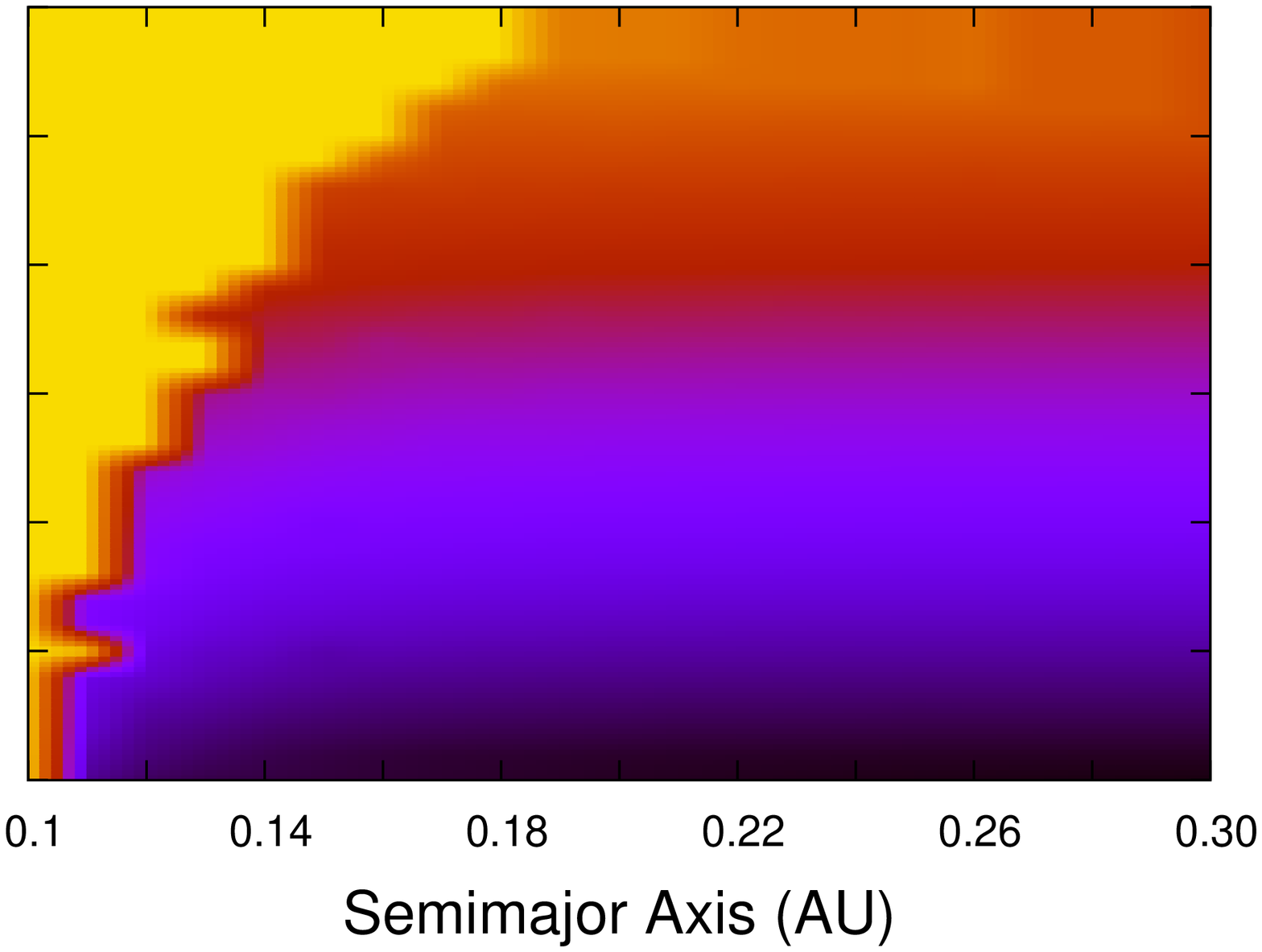}
\includegraphics[width=4.1cm,angle=270]{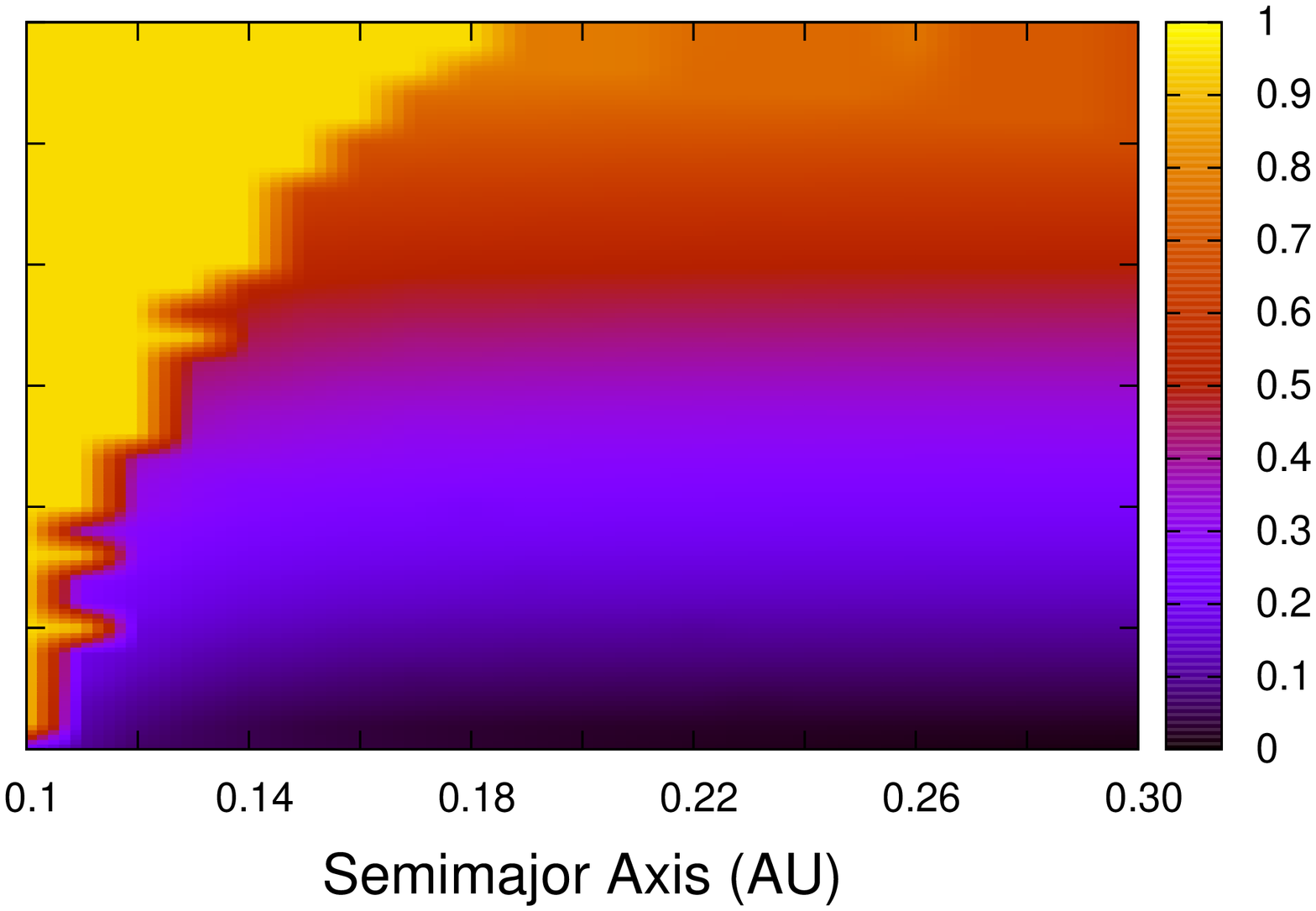}
}
\hbox{
\includegraphics[width=4.1cm,angle=270]{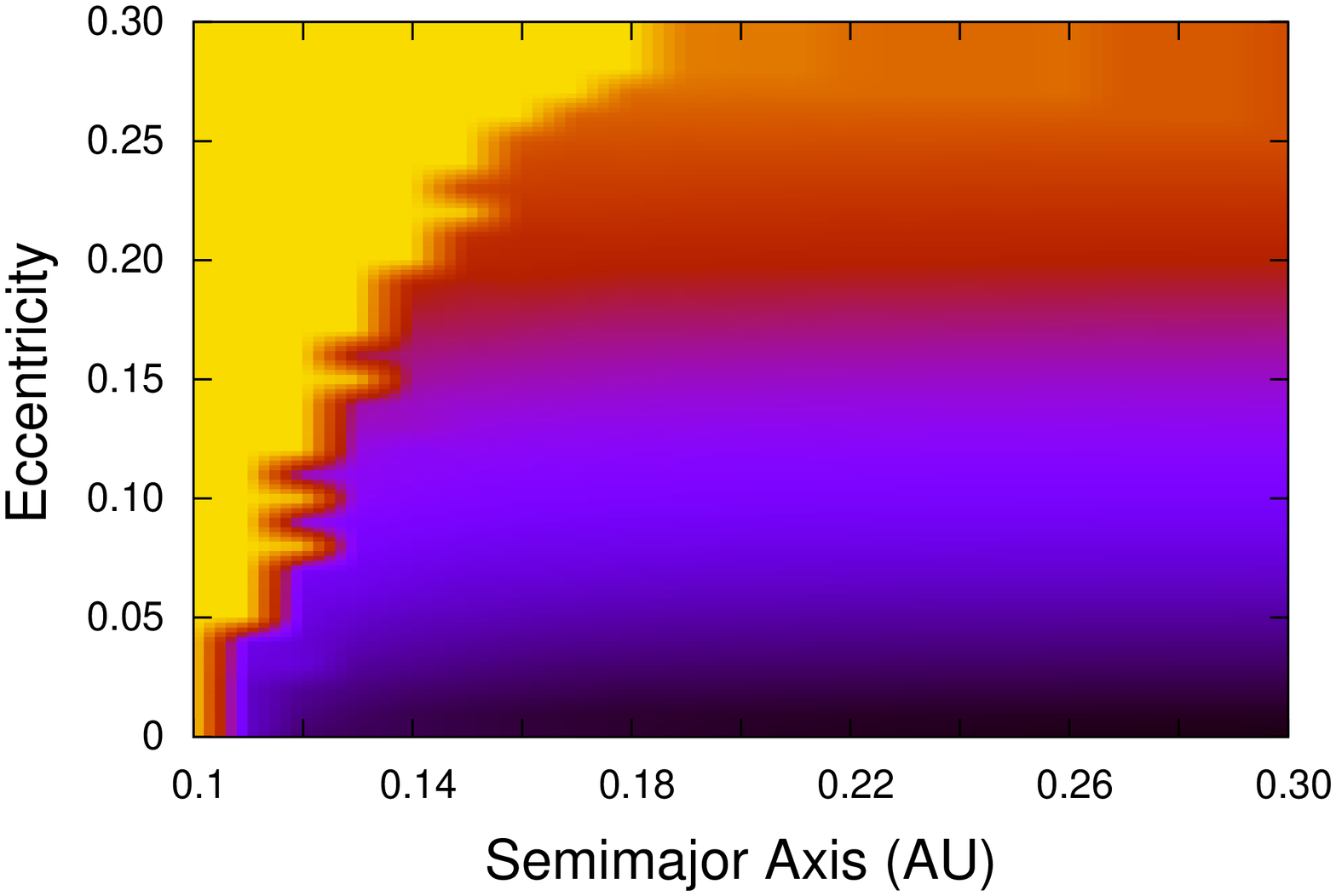}
\includegraphics[width=4.1cm,angle=270]{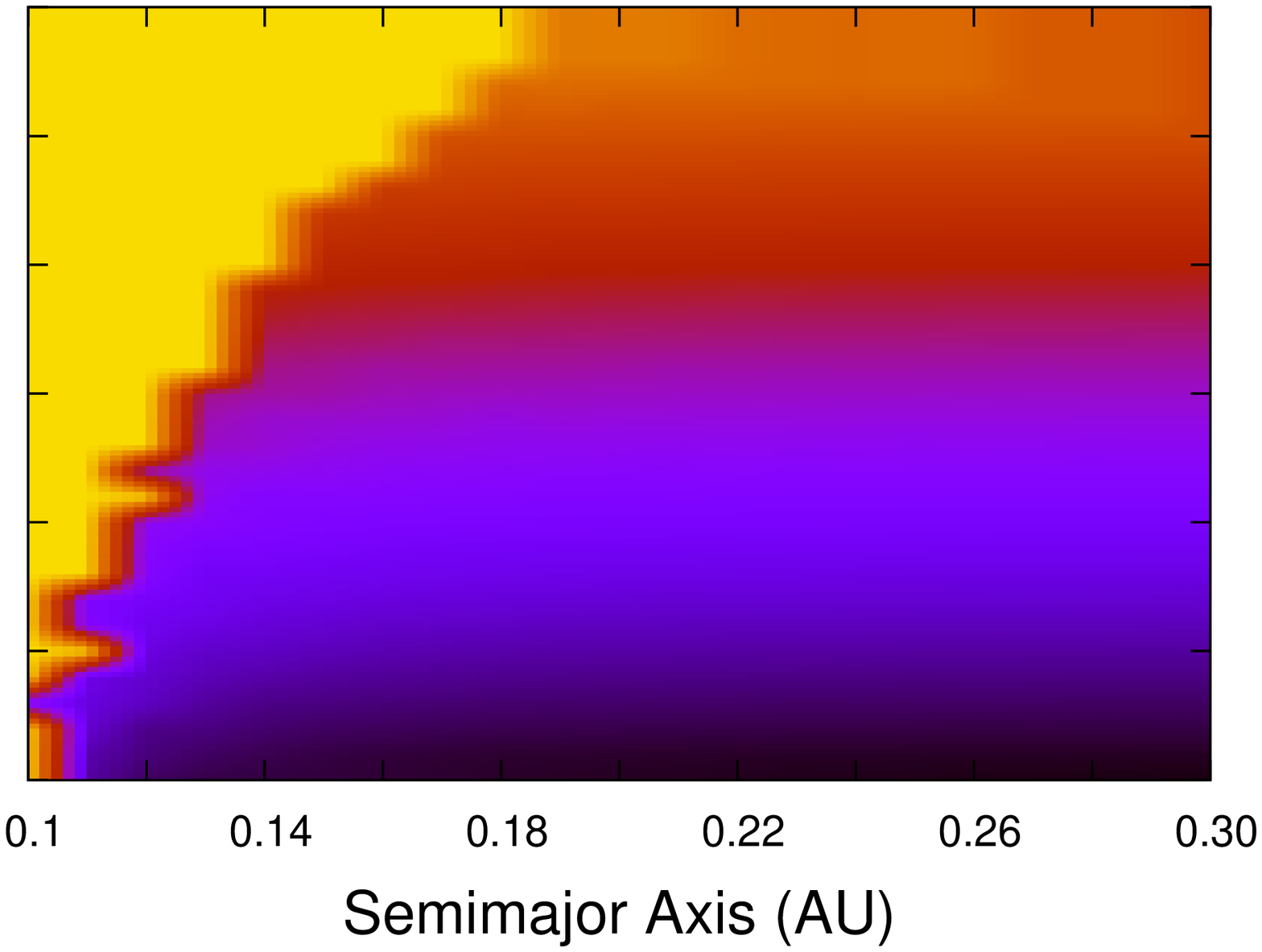}
\includegraphics[width=4.1cm,angle=270]{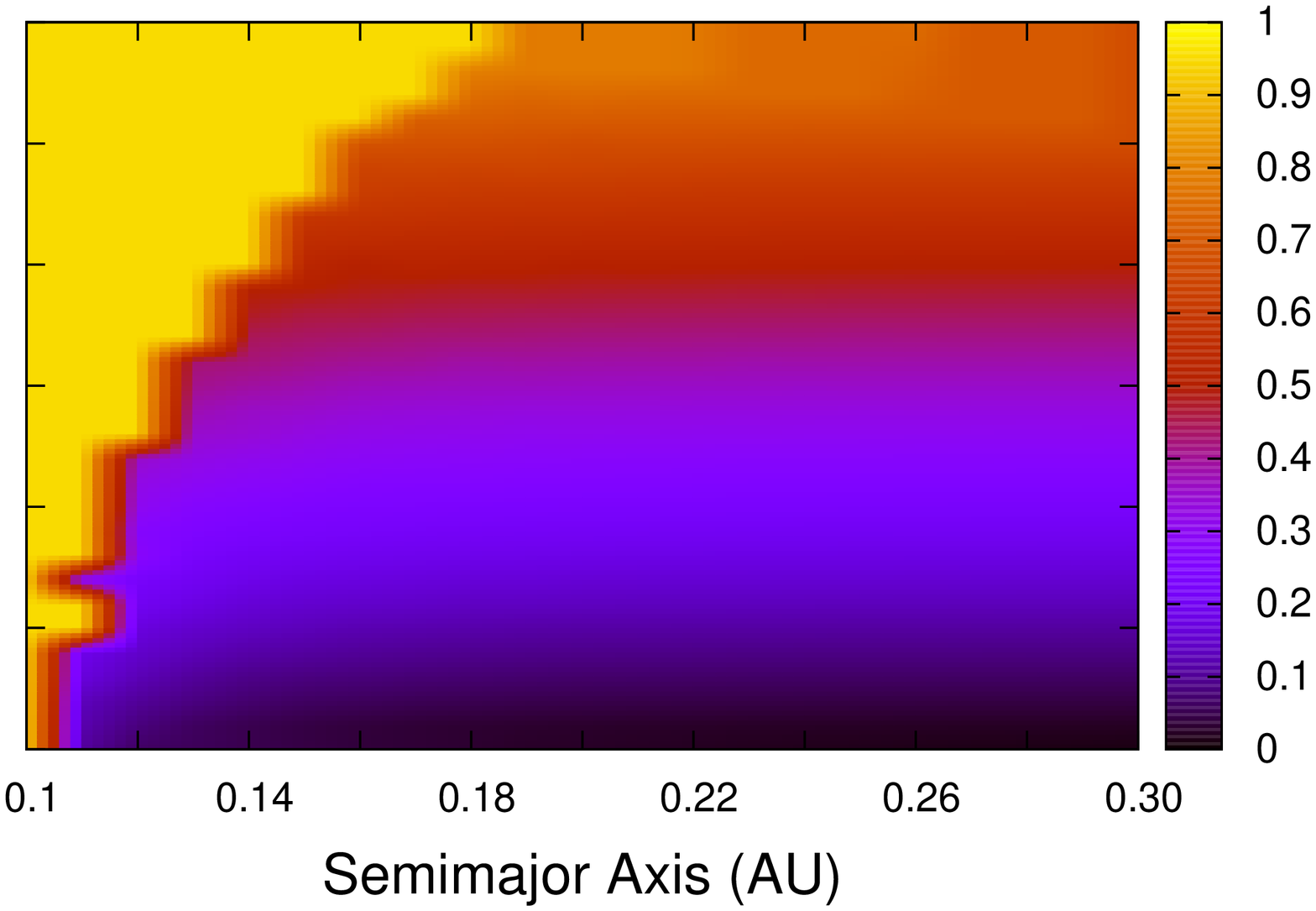}
}
%\vspace*{5cm}
\caption{The $e_{\rm max}$ maps of a circumbinary planet in the binary model 1 (left column),
2 (middle column), and 3 (right column). The top row shows the maps for a
$1\;M_J$ planet, the middle row is for a $5\;M_J$ planet, and the bottom row
corresponds to a $10\;M_J$ object. The violet and blue regions represent small values of 
$e_{{\rm max}}$ and 
correspond to bound orbits whereas the yellow region points to escape.}
\label{fig2}
\end{figure*}

\section{Stability analysis}

Prior to calculating the variations in the eclipse timing, it is
necessary to identify regions where a P-type planet can have a stable orbit.
For this purpose, we integrated the Newtonian three-body system of the binary and its 
planet for different values of the planet's mass and orbital 
elements. The mass of the planet was taken to be 1, 5, and 10
Jupiter-masses. The values of its semimajor axis and orbital eccentricity
with respect to the barycenter of the system were varied between 
$a=0.1-0.3$ AU and $e=0-0.3$ in increments of $\Delta a=0.01$ AU
and $\Delta e=0.01$, respectively.

Numerical integrations were carried out using the Lie-series method
and Bulirsch-Stoer integrator \citep{li,hans85,eggl}. 
We integrated the system for $10^4$ years which corresponds to the
following number of periods of the planet at 0.1 AU:

\begin{itemize}
\item {\bf model 1:} $2.45\times 10^5$ periods,
\item {\bf model 2:} $3.90\times 10^5$ periods,
\item {\bf model 3:} $4.50\times 10^5$ periods.
\end{itemize}

\noindent

The stability of the planetary motion was controlled by measuring the value
of its maximum orbital eccentricity ($e_{max}$). We monitored the changes in the eccentricity 
of the planet throughout the integration and determined its highest value. If the orbit of the 
planet became parabolic ($e=1$), we considered the system to be unstable.

Figure 2 shows the values of planet's $e_{\rm max}$ at different distances
from the barycenter of the binary. The left column corresponds to the binary model 1,
the middle column to binary model 2, and the right column to binary model 3. Also,
from top to bottom, each row corresponds to the values of $e_{\rm max}$ for a 1 $M_J$, 
5 $M_J$, and 10 $M_J$, respectively.  
In each panel, the axes represent the initial
values of the semimajor axis and eccentricity of the planet at the beginning
of the integration. The color at each point depicts
the maximum value that the eccentricity of the planet acquired during the integration.
As indicated by the scale on the right side of the figure, yellow 
corresponds to parabolic orbits and denotes instability.

An inspection of the $e_{\rm max}$ maps shown in Fig. 2 indicates that
for planets in circular or low eccentricity orbits, the range of 
the semimajor axis for which the planetary orbit may be bound is large and does not 
change much for different values of the planet's mass. For instance, as shown
by the top left panel of this figure, the boundary of the unstable zone for a 1 $M_J$ 
planet in a circular orbit in the binary model 1 is interior to 0.1 AU. This suggests 
that all circular orbits with semimajor axes larger than 0.1 AU in this model may be bound. 
As the mass of the planet grows (middle and bottom panels of the left column in Fig. 2),
the outer boundary of the unstable region slowly progresses toward larger semimajor axes. Such a trend
is also seen in the $e_{\rm max}$ maps of binary models 2 and 3. 
As shown by the lower panels in the middle and right columns, the boundary of the unstable zone 
for circular orbits is slightly shifted outward to 0.11 AU.
These results are consistent with the results of the stability analysis of a test particle in
a circumbinary orbit as presented by Dvorak et al. (1989) and Holman \& Wiegert (1999).
According to the estimate of the boundary of the stable zone as given by these authors, 
the critical distance beyond which the orbit of a Jovian planet in all our three binary 
models will be stable is in the range of 0.1 to  0.12 AU.

While the value of $e_{\rm max}$ can be used to identify parabolic (unstable) orbits, it cannot be
used as a rigorous indicator of planet's orbital stability. A low value of $e_{\rm max}$ implies 
a bound planetary orbit. But that is only for the duration of the integration, and there is 
no guarantee 
that the orbit of the planet will stay bound for a long time. In order to determine the orbital
stability of the planet, we used the Fast Lyapunov Indicator (FLI) \citep{froe97}.
The FLI is a chaos indicator that measures the exponential divergence 
of nearby trajectories and distinguishes between regular and chaotic motion. 
For details of this technique, we refer the reader to \cite{froe97}, \cite{lega} 
and \cite{fouchard}. Applications of FLI to planetary motion in binary star systems 
can be found in \cite{loh02}, \cite{loh03} and \cite{haghighi}.

\begin{figure}
\includegraphics[width=5cm,angle=270]{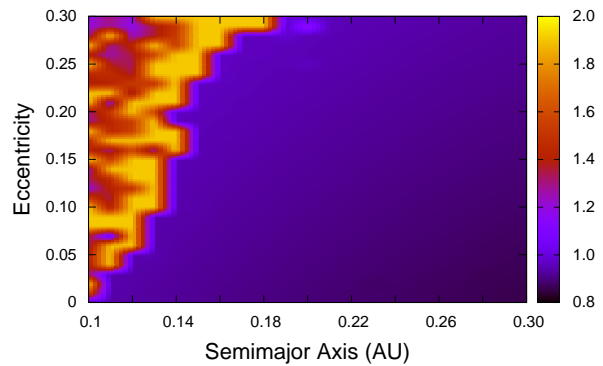}
\caption{The FLI stability map for a 1 $M_J$ P-type planet in the binary model 1.
The violet and blue regions (small values of the FLI) correspond to regions of regular 
motion, and red and yellow denote chaotic orbits.}
\label{fig3}
\end{figure}

Figure \ref{fig3} shows a sample of an FLI stability map for a 1 $M_J$ planet in 
binary model 1. The colors in this figure represent the values of FLI (the logarithmic scale 
on the right) which depict the degree of chaos. Dark colors correspond to less chaotic and  
regular orbits. As shown by this figure, there is a region in the parameter-space where the
orbit of the planet is likely stable (dark area between 0.1 and 0.3 AU). 

A comparison between the dark region of Fig. 3 and its corresponding region in the system's 
$e_{\rm max}$ map in Fig. 2 (upper left panel) points to an interesting observation: 
even though the maximum values of the planet's orbital eccentricity can reach to high values, 
the orbit of the planet may still be stable. In other words,
elevated planetary eccentricities do not necessarily correspond to instability. 
For instance, as shown by the upper left panel of Fig. 2, orbits with initial eccentricities 
of 0.3 may have $e_{max}$ values up to 0.7. Nevertheless, according to the FLI results shown 
in  Fig. \ref{fig3}, they are probably long-term stable.

Similar comparison can also be made between the chaotic region of Fig. 3 (indicated by red and
yellow colors) and the system's $e_{\rm max}$ map in Fig. 2. The results point to an inverse
phenomenon, that is, instability among seemingly bound orbits. For instance, while for semimajor 
axes ranging from $a=0.11$ AU to $a=0.14$ AU, and the eccentricities between $e=0.05$ and $e=0.15$, 
the $e_{max}$ map suggests bound planetary orbits, the FLI stability map indicates chaotic motion. 
The different color shades for chaotic orbits in the FLI map have resulted from the fact 
that the degree of divergence depends on the initial conditions of the orbits.
Even though this shading may suggest less chaotic 'islands' in parameter-space, 
when compared to the corresponding $e_{\rm max}$ map, one can see that orbits inside 
the chaotic border zone become parabolic.

Using the combination of $e_{\rm max}$ and FLI analyses, we identified the stable planetary orbits 
in all our three binary models. As an example, the number of stable circular orbits are shown
in Table \ref{tab2}. As expected, stability is almost independent of the mass-ratio of the binary.

We also  carried out simulations for non-zero values of
the binary's eccentricity. Figure 4  shows the maps of $e_{\rm max}$ for a 1 $M_J$ planet
in the binary model 1. The binary eccentricity was chosen to be $e_{\rm bin}=$ 0.05, 0.1, and 0.15. 
Table \ref{tab3} shows the number of stable orbits for these simulations. 
As shown here, the unstable
region expands out by more than two folds when $e_{\rm bin}=0.1$. Our analysis indicated that
for the value of the binary eccentricity $e_{\rm bin}=0.2$, only 4 orbits remained stable (Table 2).

\begin{figure}
\includegraphics[width=5cm,angle=270]{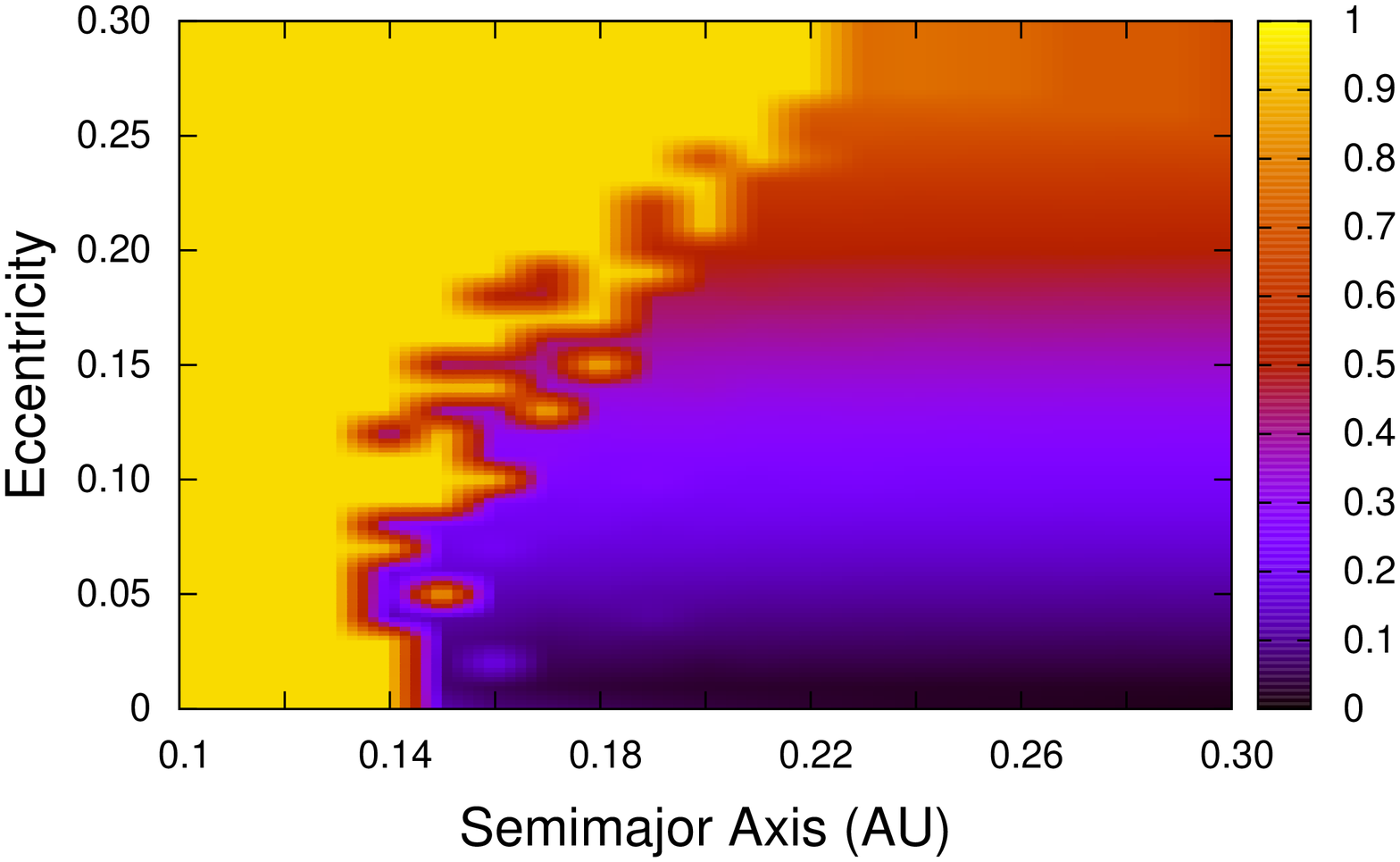}
\includegraphics[width=5cm,angle=270]{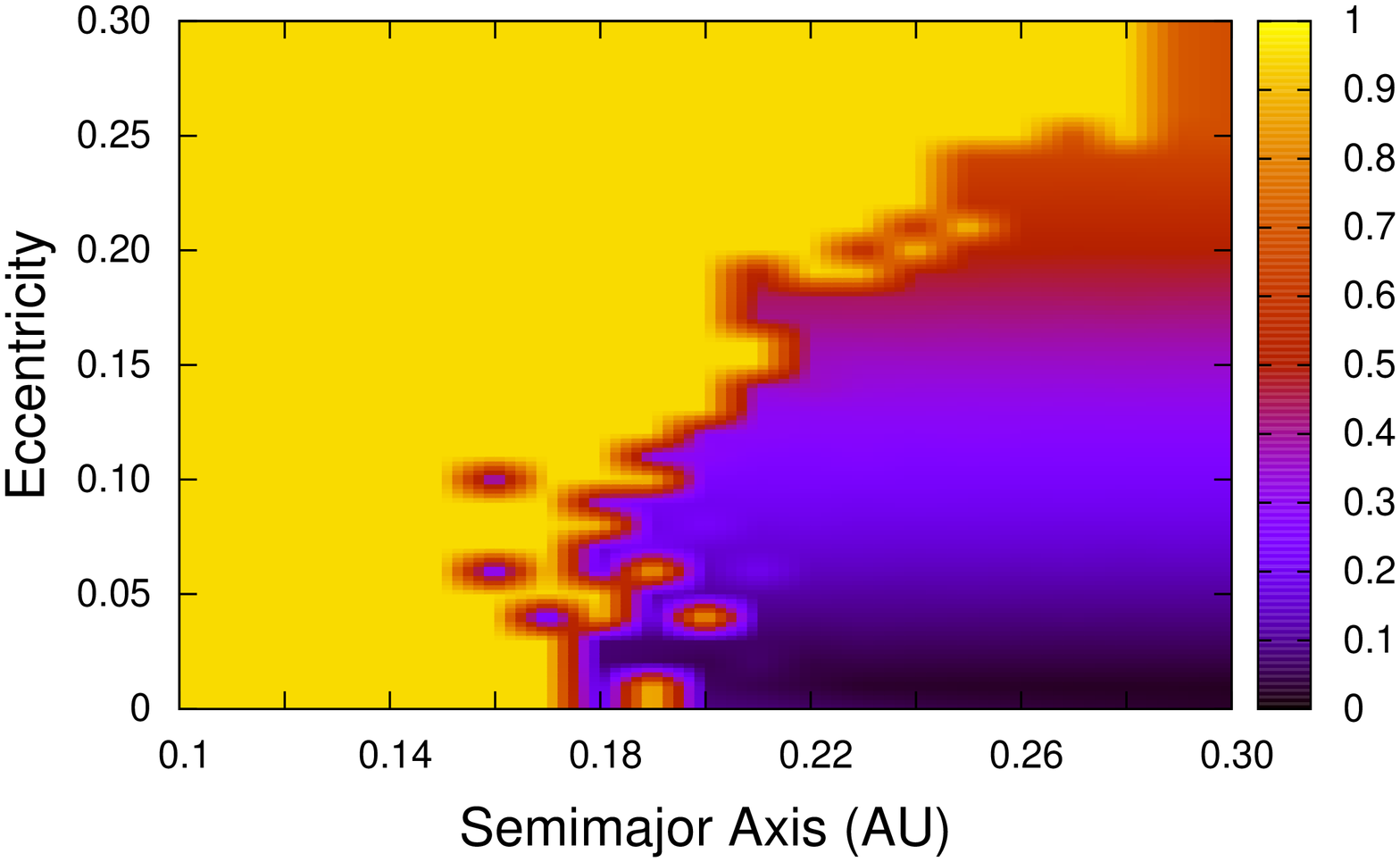}
\includegraphics[width=5cm,angle=270]{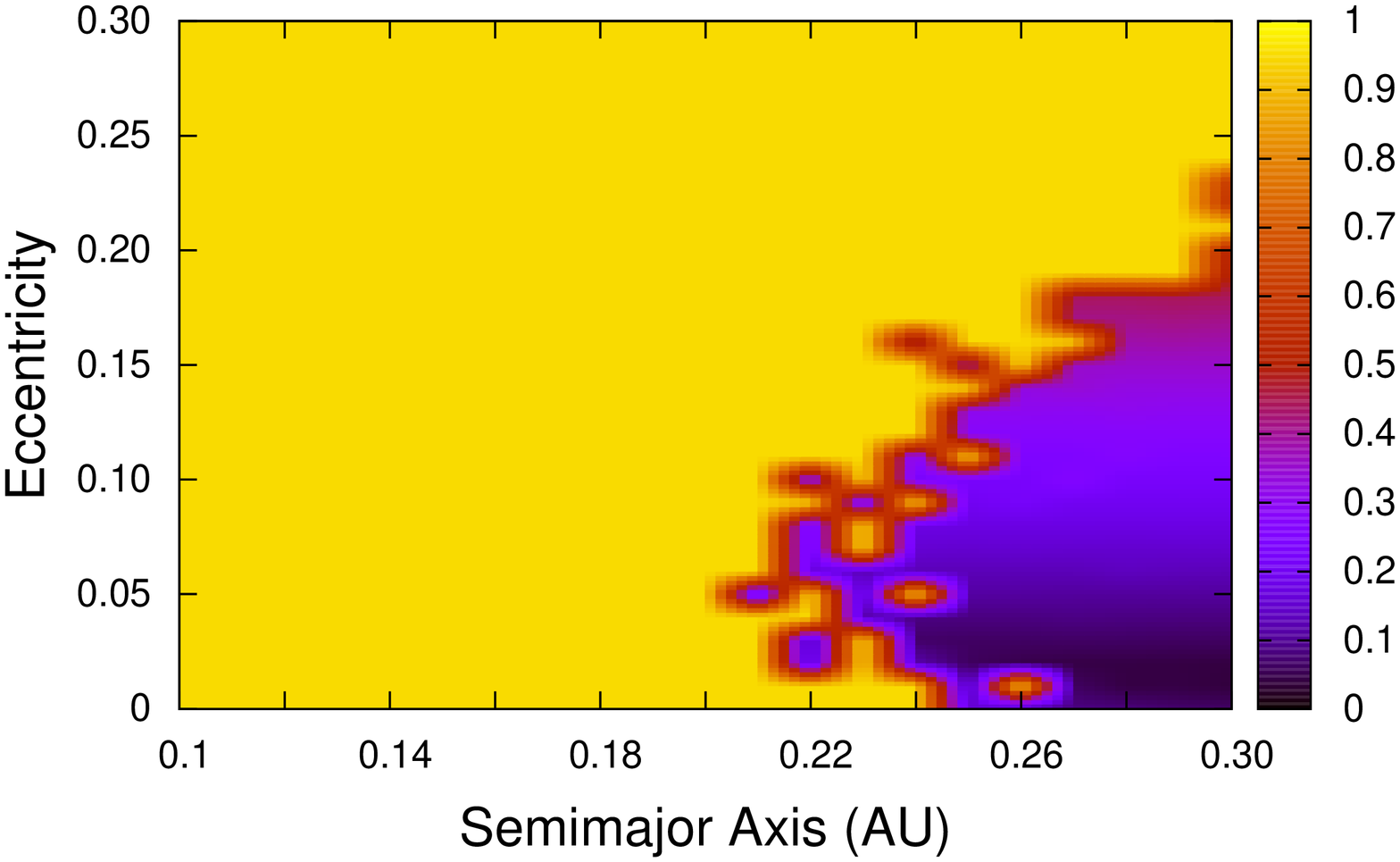}
\caption{The $e_{\rm max}$ maps  of a 1 $M_J$ planet in the binary model 1 with 
$e_{\rm bin}=0.05$ (top), 0.1 (middle), and 0.15 (bottom).}
\label{fig4}
\end{figure}

The results of our stability analysis allow us to focus our calculations of ETVs
on the region of the parameter-space where P-type planets have stable orbits. 
As shown, the planet's stable region shrinks for 
large values of its eccentricity as well as the eccentricity of the binary. Initial
binary eccentricities beyond $e_{\rm bin}=0.15$ move this region to large distances ($a=0.24$ AU) 
where the perturbing effect of the 
planet on the dynamics of the binary becomes negligible (Fig. \ref{fig4}). 
The closest possible stable orbit for a planet 
($a=0.1$ AU) is when the planet is Jupiter-mass, and both the planet and binary have 
circular orbits.
As a result, for the purpose of calculating ETVs, we only consider fully circularized binaries. 

\begin{table}
\centering
 \caption{Number of stable orbits of all three binary models for $e_{\rm bin}=0$.}
  \begin{tabular}{lccc}
  \hline
Model & $M_{\rm planet}$ & stable orbits  & stable orbits\\
      & ($M_J$)     & (number)    &  ($\%$)       \\
  \hline
{\bf model 1}&1   &   540  &   82.9  \\
&5   &   530  &   81.4  \\
&10  &   527  &   80.9  \\
  \hline
{\bf model 2}&1   &   531  &   81.5  \\
&5   &   529  &   81.2  \\
&10  &   526  &   80.7  \\
  \hline
{\bf model 3}&1   &   539  &   82.6  \\
&5   &   538  &   82.6  \\
&10  &   533  &   81.8  \\
\hline
\end{tabular}
\label{tab2}
\end{table}
    
\begin{table}
\centering
 \caption{Number of stable orbits for binary model 1 and for different
values of $e_{\rm bin}$.}
  \begin{tabular}{lccc}
  \hline
$e_{\rm bin}$ & $M_{\rm planet}$ & stable orbits  & stable orbits\\
      & ($M_J$)            & (numbers)         & ($\%$)        \\
  \hline
0.00 &1   &   540  &   82.9  \\
0.05 &1   &   407  &   62.5  \\
0.10&1   &   270  &   41.4  \\
0.15&1  &   127  &   19.5  \\
0.20&1  &    4  &   0.6  \\
\hline
\end{tabular}
\label{tab3}
\end{table}

\section{Calculation of Eclipse Timing Variations}

As indicated by the results of the stability analysis, a Jovian
planet in a P-type orbit may be stable in the vicinity of a binary star system.
Although small, the gravitational perturbation of this planet may 
affect the motions of the two stars and cause their orbits to deviate from 
Keplerian. In an eclipsing binary, these deviations result in
variations in the time and duration of the eclipse. 
Similar to the variations in the transit timing of a planet 
due to a second perturber \citep{mira,hol,agol07,kip09a,kip09b,ford}, the variations 
in eclipse timing can be used to infer 
information about the mass and orbital elements of the circumbinary planet. In this 
section, we calculate the eclipse timing variations of our model binary star systems
for different values of the mass and orbital parameters of the binary and its
P-type planet. Our goal is to identify a range of these parameters for which
the magnitude of ETVs will be within the temporal sensitivity of CoRoT and
{\it Kepler} space telescopes.

We simulated the dynamics of our binary models and their P-type planets 
in a barycentric coordinate system.
Simulations were carried out for 1 year corresponding to 
approximately 70 transits of the secondary star. We calculated ETVs
by determining the difference between the eclipse 
timing $(t_1)$ of the unperturbed system (star-star) and its corresponding value 
$(t_2)$ in the perturbed case (star-star-planet)\footnote{We chose to integrate 
the binary two body problem instead of taking 
analytical solutions in order to minimize
the influence of the numerical integration algorithm on the results.}.
Additionally, we subtracted the planetary induced constant rate of apsidal
precession. We also Fourier-analyzed the resulting signal and 
compared the superposition of the main frequencies to simple estimates 
on maximum amplitudes 
$[(dt_{max}-dt_{min})/2]$ for the entire duration of integration. 
In cases where the differences between Fourier-composition-derived ETVs and maximum 
amplitudes were lower than $10$\%, we chose to present the ETVs' maximum amplitudes.

\subsection {Implication for the detection of circumbinary planets}

As mentioned earlier, we would like to identify regions of the parameter-space
for which the ETV signals of an eclipsing binary will be detectable by
CoRoT and {\it Kepler} space telescopes.
Since as indicated by the stability analysis, stable planetary orbits move
to large distances as the eccentricity of the binary increases, we limited our
calculations to circular systems. 
We recall that in all our models, the separation
of the binary is 0.05 AU.

For the planet's motion, we considered both resonant and non-resonant orbits.
Similar to transit timing variation, ETV signals are strongly enhanced when the P-type 
planet and the binary are in a mean-motion resonance (MMR). Results of our stability analysis
indicate that the location of the 2:1 MMR at 0.079 AU is too close to the binary's center of 
mass to be stable (Dvorak et al. 1989, Holman \& Wiegert 1999). We, therefore, considered 
cases where the planet is in a 3:1 (0.104 AU) and a 4:1 (0.126 AU) MMR. 

For the non-resonant orbits,
we chose the semimajor axis of the planet to have the values of 0.2 AU and 0.3 AU.
The eccentricity of the planet's orbit in all cases was chosen to be 0 or 0.15,
and its angular variables (inclination, mean-anomaly, longitude of the ascending node,
argument of pericenter) were set to zero.

\begin{figure}
\includegraphics[width=9cm,angle=0]{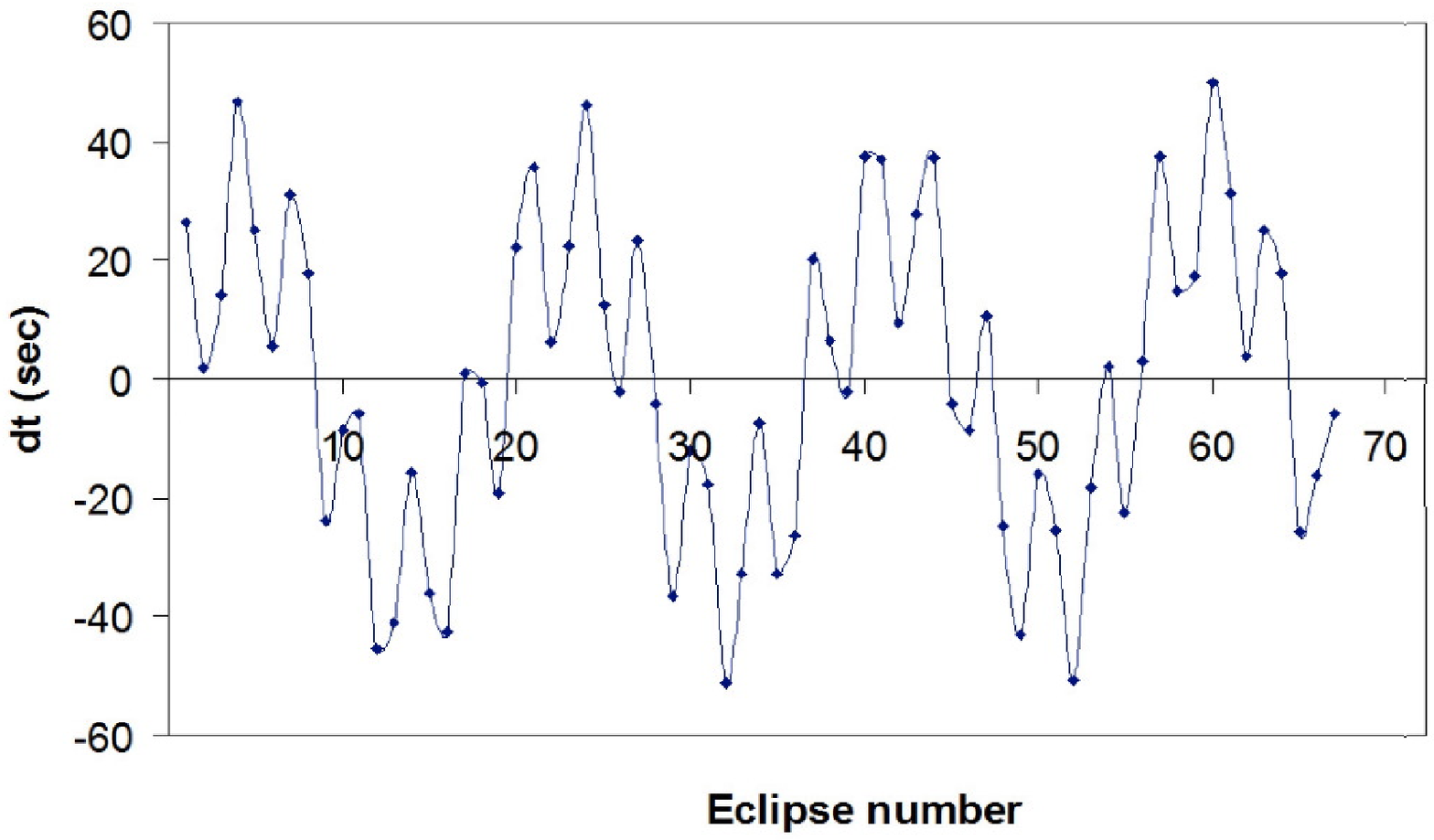}
\vskip -90pt
\includegraphics[width=9cm,angle=0]{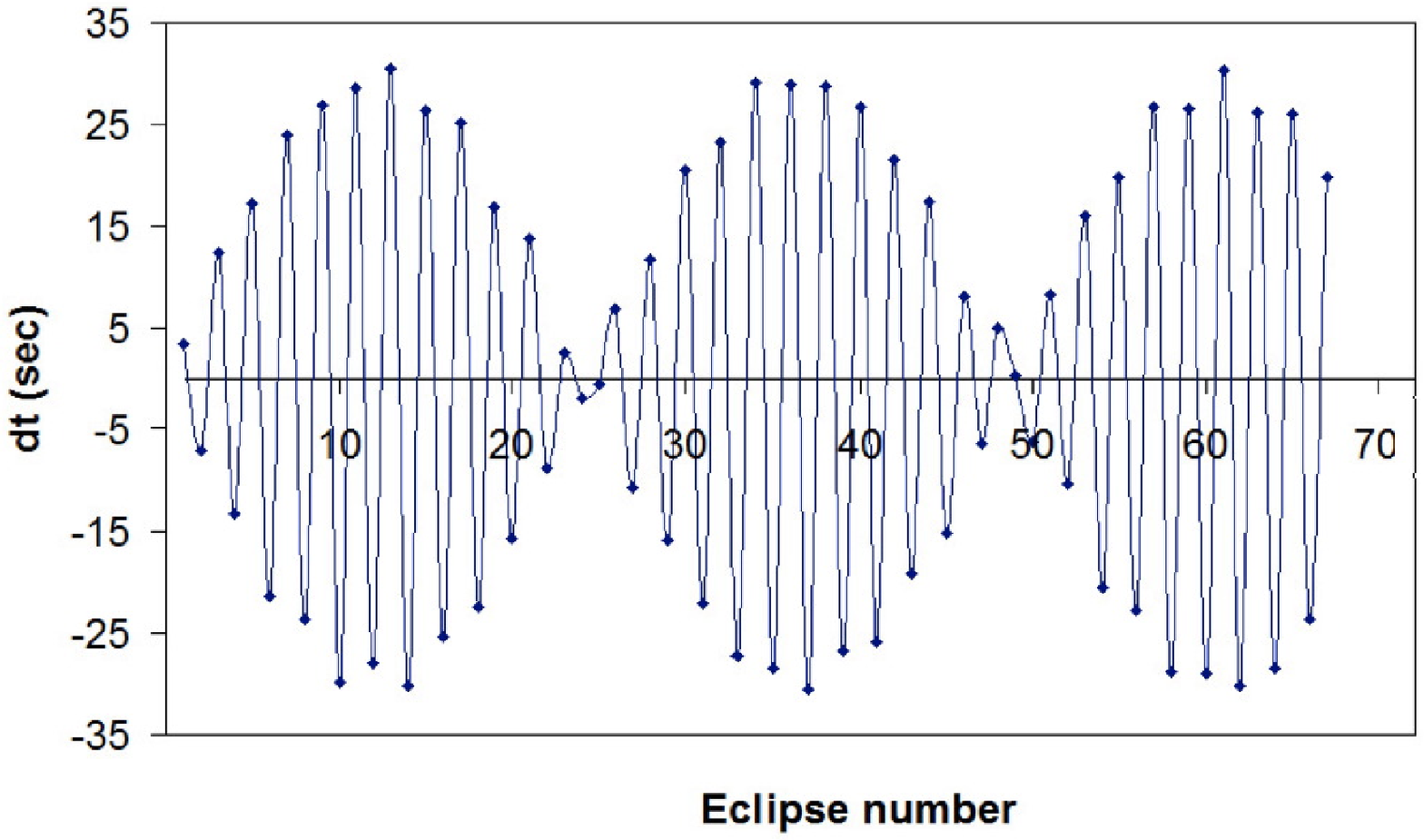}
\caption{The graphs of ETVs for a 1 $M_J$ planet in a circular orbit in
the binary model 3. The planet is in a 3:1 MMR in the top panel and a 4:1 MMR
in the bottom one.}
\label{fig5}
\end{figure}

Table 3 shows the results of the calculation of ETVs for a circular binary.
Even though MMRs did not have too much influence on the stability of the system, 
the most prominent ETVs were found at 3:1 and 4:1 
resonances. Figure \ref{fig5} shows these ETVs for a 1 Jupiter-mass planet.

In order to assess the detectability of our ETV signals with CoRoT and {\it Kepler},
we compared our results with the values of the Detectable Timing Amplitudes (DTA) of 
these telescopes. As shown by \citet{syb}, 
given the typical photometric error, the value of the DTA for 
CoRoT is approximately 4 sec for a 12 mag star and 16 sec for a star with 15.5 mag.
Similar analysis suggests that stars with magnitudes between 9 and 14.5 mag 
will have DTAs ranging from 0.5 to 4 sec with {\it Kepler}.

\begin{figure*}
\hbox{
\includegraphics[width=5.99cm,angle=0]{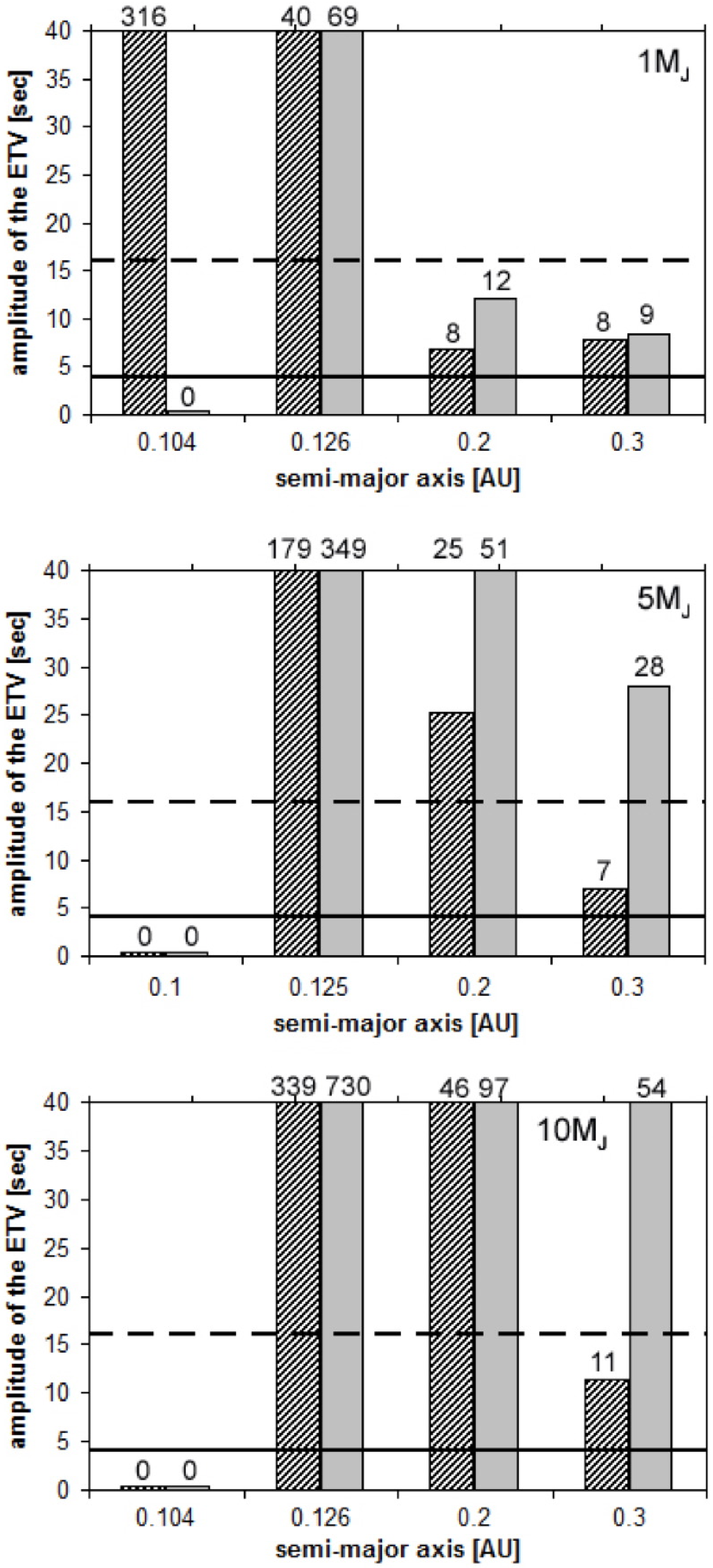}
\includegraphics[width=5.32cm,angle=0]{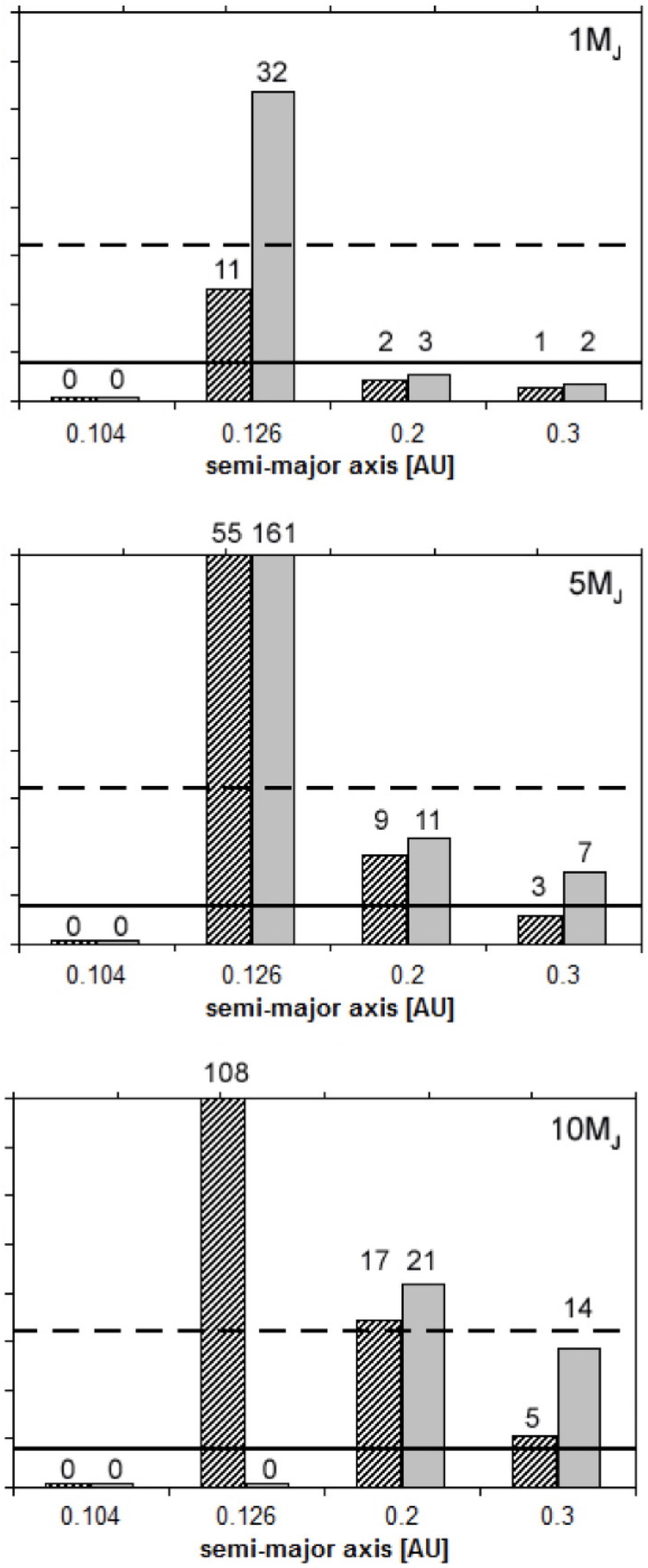}
\includegraphics[width=5.33cm,angle=0]{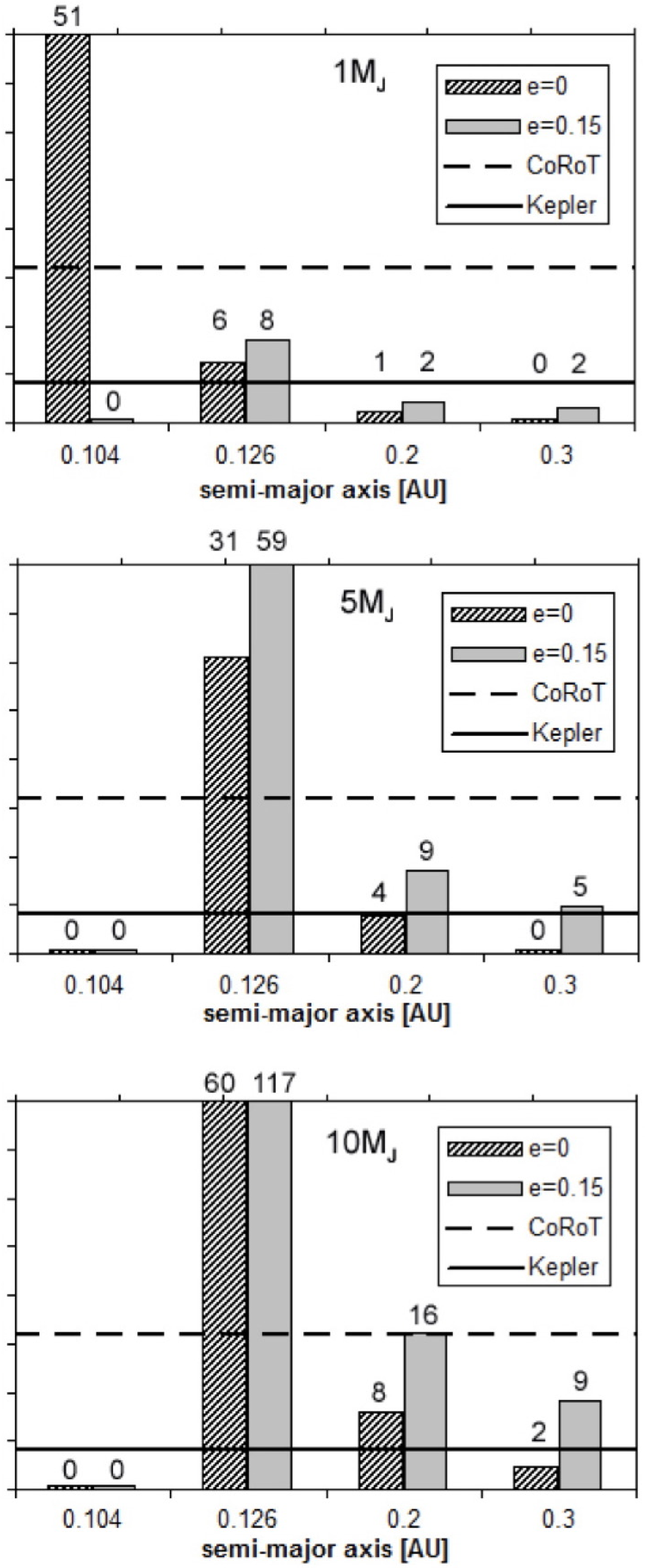}
}
\caption{Graphs of the maximum amplitude of ETVs for all three binary models with $e_{\rm bin}=0$.
The locations of bars on the horizontal axis correspond to 3:1 MMR at 0.104 AU, 4:1 MMR
at 0.126 AU, and two non-resonant orbits at 0.2 and 0.3 AU. Graphs are shown for two
values of the planet's orbital eccentricity (0 and 0.15). As a point of comparison, the maximum values 
of DTAs for CoRoT and {\it Kepler} \citep{syb} are also shown.}
\label{fig6}
\end{figure*}

Figure \ref{fig6} shows a comparison between the values 
of DTA calculated by \citet{syb} and the amplitudes of ETVs obtained from our simulations.
The x-axes in this figure represent the planet's initial semimajor axis and the y-axes show 
the maximum amplitude of ETVs. In cases where the value of ETV is larger than 40 sec, this value 
is indicated on the top of its corresponding bar. As a point of comparison,
the maximum values of the DTAs for CoRoT (16 sec for a 15.5 mag star) and {\it Kepler} 
(4 sec for a 14.5 mag star) are also shown. When the ETV signal is higher than these observational 
thresholds, we assume that a planet may be indirectly detectable via eclipse timing 
of the secondary star.

As shown in this figure, planets in mean-motion resonances present the highest
probability of detection. This is expected as these planets create the largest
variations in eclipse timing. Also, as indicated by the left column, binary 
model 1 with low-mass stars present the best prospect for detecting P-type
Jovian planets. As the masses of the binary stars increase, the prospect
of detection shifts toward larger planets. For instance, in binary model 3 (right column), 
a 5 Jupiter-mass planet can be detected in orbits smaller than 0.2 AU. As expected, 
the highest prospect of detection is for a 10 Jupiter-mass object (lower right panel).
Figure 6 also shows that while planets in circular orbits produce the largest ETVs 
when in resonance, slight eccentricity in the orbit of the planet increases
the prospect of its detection and extends its detectability to large distances.
The latter is more pronounced for larger planets around more massive binaries
as in these cases, the planetary orbit stays stable when its orbital eccentricity
is slightly increased.

\begin{table}
\centering
 \caption{Systems with ETVs that are potentially detectable with CoRoT and {\it Kepler}.}
  \begin{tabular}{lllll}
  \hline
 Model       &$M_{\rm {planet}}$ & $\>\>\>\,a$ & $\>\>\>\,e$ & ETV\\
             &$\>\>\,(M_J)$   &  (AU)   &   & (sec) \\ 
  \hline
{\bf model 1}& 1  & 0.104   &  0  & 316\\
             & 1  & 0.126   &  0, 0.15 & 40, 69\\
             & 5  & 0.126   &  0, 0.15 & 179, 349\\
             & 5  & 0.2     &  0, 0.15 & 25, 51\\
             & 5  & 0.3     &  0.15 & 28\\
             & 10 & 0.126   &  0,0.15 & 339, 730\\
             & 10 & 0.2   &  0, 0.15 & 46, 97\\
             & 10 & 0.3     &  0.15 & 54\\
\hline
{\bf model 2}& 1  & 0.126   &  0.15 & 32\\
             & 5  & 0.126   &  0, 0.15 & 55, 161\\
             & 10  & 0.126   &  0   & 108\\
             & 10 & 0.2     &  0, 0.15 & 17, 21 \\
\hline
{\bf model 3}& 1  & 0.104   &  0 & 51 \\
             & 5  & 0.126   &  0, 0.15 & 31, 59\\
             & 10  & 0.126   &  0, 0.15  & 60, 117\\
             & 10 & 0.2     &  0.15 & 16\\
\hline
\end{tabular}
\label{tab5}
\end{table}

\section{Discussion}

We have studied the prospects of the detection of circumbinary planets with
CoRoT and {\it Kepler} space telescopes using the variations of binary eclipse timing. 
The uninterrupted high precision photometry of these telescopes 
\citep{Koch04, Alonso08} has given them 
unique capability for detecting small variations in eclipse timing measurements. 
We have calculated such variations in several binary models with circumbinary planets
and compared the results with the temporal sensitivity of these telescopes. 
As expected, the prospect of detection is higher for planets in resonant orbits
as these objects create larger ETVs.  This is more pronounced when the stars of the 
binary have low masses.
This result is consistent with the findings by \cite{schneiderd}. However, it is 
important to note that low-mass binaries may be hard to identify.

Our study indicates that around solar-mass binaries, planets with sizes down to 1 Jupiter-mass 
can produce detectable signals. In this case, planets
have to move in almost circular orbit close $(a \sim 2\,a_{\rm {bin}})$ to binary's center of mass
in order to maintain stability.  
More massive planets on higher eccentricities may orbit the binary at 
distances up to approximately $6\,a_{\rm {bin}}$ and still be able to 
produce ETVs that are detectable by {\it Kepler}.

Result of our simulations indicate that
although the orbital stability of the planet is strongly affected by increasing 
its eccentricity and the eccentricity of the binary, slight deviations
from circular orbits, in particular in the orbit of the planet, result in its
periodic close approaches to the binary and creating large ETVs. Our study
suggests that when not in a resonance, in general, giant planets on slightly 
eccentric circumbinary orbits show bigger prospects for having detectable ETVs. 
With their high precision photometry and long  duration of operation, CoRoT 
and {\it Kepler} are well suited for indirect planet detection via ETVs, 
and have the capability of detecting such planets within the durations of their operation.

\section*{Acknowledgments}
RS acknowledges supports through the MOEL grant of the \"OFG (Project MOEL 386) and 
the FWF project P18930. NH acknowledges support from the NASA Astrobiology Institute under 
Cooperative Agreement NNA04CC08A at the Institute for Astronomy, University
of Hawaii, and NASA EXOB grant NNX09AN05G. Supports are also acknowledged through 
the Austrian FWF Erwin Schr\"odinger grant no. J2892-N16 for BF, the Austrian FWF project no. P20216
for SE, and the Austrian FWF project no. P22603, P20216 for E. P-L. 
We thank J. Schneider and G. Wuchterl for fruitful discussions.

\end{document}